# Magical or magnetic? Less commonly taught facts about real-world permanent magnets and their diverse interactions with objects

Dr. Yan-Cong Chen[1*]

[1]Key Lab of Bioinorganic and Synthetic Chemistry of Ministry of Education, School of Chemistry, Sun Yat-Sen University, Guangzhou 510006, China.

**Permanent magnets are fascinating. They generate magnetic field and act as the key component in compasses, motors, speakers for practical applications. They are also made into magnetic toys for entertainment and education with magical magnetic tricks. Some of these tricks fueled the recent upsurge for magnetic-responsive objects including but not limited to a LK-99 mixture.[1-2] A lack of full consideration of the magnetic interactions seems to be the reason why some tricks like "half levitation" is misinterpreted, and the situation is not really rectified in the following replications and discussions.[3-5] Here I would like to go through some less commonly taught facts about real-world permanent magnets starting from the nonuniformity to their combined effects with other factors. Based on these facts, I will also break down the situation of some examples of how permanent magnets diversely interact even with ordinary objects, as one may have seen from recent sources.[2-5] I believe this discussion is useful both in an informative and educational way, helping those who want to think deeper about magnetic mechanism beyond a simple attract/repel terminology.**

*E-mail: chenyc26@mail.sysu.edu.cn

## Fact #1: The magnetic field near the surface of a permanent magnet is highly nonuniform .

At a distance, a permanent magnet may be regarded as a simple N-S dipole model with no fine structure at all. A simple approximation of "the larger the distance, the weaker the magnetic field" almost tells the entire story. However, when one is talking about the "magic tricks" near its surface, the spatial distribution of the magnetic field of a real-world permanent magnet plays a crucial role. Here I measured the surface magnetic fields at some representative points of some of my NdFeB magnets using a tesla meter and report the results in a simplified (axial, radial) way:

Cylinder Magnet No.1 ($\varphi$ = 12 mm, $h$ = 8 mm, Fig. 1a): (400 mT, 10 mT) at the middle 1A, (290mT, 380mT) near the edge 1B, and (-150mT, 5mT) aside 1C.

Cylinder Magnet No.2 ($\varphi$ = 20 mm, $h$ = 10 mm, Fig. 1b): (400 mT, 2 mT) at the middle 2A, (420 mT, 110 mT) at the half-radius 2B, (200mT, 460mT) near the edge 2C, and (-220mT, 3mT) aside 2D.

Cuboid Magnet No.3 ($l$ = $d$ = 29 mm, $h$ = 19 mm, Fig. 1c): (130 mT, 2 mT) at the middle 3A, (159 mT, 28 mT) at the half-way 3B, (230mT, 170mT) near the edge 3C, and (-150mT, 2mT) aside 3D.

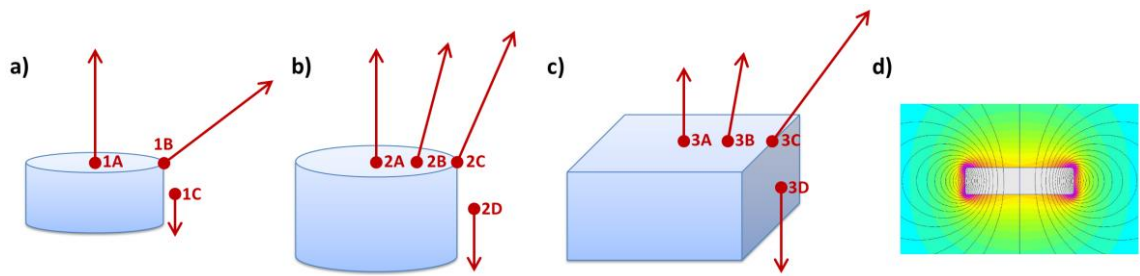

**Figure 1.** Experimental magnetic fields (a-c, red arrows) for some of my NdFeB permanent magnets with the N pole on the upper surface, along with (d) the simulation for a D82 disc magnet.[6] The drawings are not to scale between different sub-figures.

Generally speaking, a piece of real-world permanent magnet in fact shows strongest magnetic field at the edge and pointing outwards, rather than at the center of its pole. In a figurative way, one can imagine that the magnet is cut into many pieces. Now the center piece has to suffer from all the (opposite) magnetic fields generated by those around it, so the final field is smaller. It should be noted that the exact spatial distribution of the magnetic field near the surface of a permanent magnet is different from pieces to pieces, and a simulation of a D82 disc magnet is also presented (Fig. 1d).[6] Sometimes the poles are not even located along an assumed axis or face, especially when one cannot tell the poles from Internet videos.

## Fact #2: A magnet field itself does not attract/repel an object, but the gradient of a magnetic field does.

A magnet can attract/repel an object, which is true. A magnetic field, however, does not attract/repel an object in the way one might expect. Mathematically, for a small object having a magnetic moment $m$ placed in a magnetic field $B$, the magnetic force is: $F_{mag} = \nabla (m \cdot B)$.[7] It is the gradient operator $\nabla$ (*Friendly reminder: it contains those $\partial/\partial x$, $\partial/\partial y$ and $\partial/\partial z$ in the Cartesian coordinates.) that generates the magnetic force, and thus the attraction/repletion.

In fact, in a perfect uniform magnet field where there is no gradient for (***m*** • ***B***), an object gets no force, no matter it is ferromagnetic/paramagnetic/diamagnetic/etc. One can also understand it from the energy point of view: The potential energy here is written as $U = -(\mathbf{m} \cdot \mathbf{B})$. If moving the object around does not change (***m*** • ***B***), it does not change $U$, so the potential energy has no gradient in all direction, so there is no force at all. This is fundamentally different from the situations in an electric field, because an electrical object can be a monopole, but a magnetic object is a dipole. Even if one imagines that a magnetic object can be separated into two magnetic monopoles, their hypothetical forces are completely cancelled out.

What about a nonuniform magnetic field like those near the surface of a permanent magnet in Fact #1? It is called nonuniform because it varies from place to place, which means gradient! A magnet can attract/repel an object because its magnetic field is nonuniform. A ferromagnetic/paramagnetic object with ***m*** and ***B*** in the same direction gets a positive attracting force to increase (***m*** • ***B***) toward the positive direction (0 → positive), so it tend to move closer to a permanent magnet for a larger ***B*** and stay near a large ***B*** area. In real world, it is not necessarily at the exact ***B***$_{max}$ point, because the shape and size of the object of the object must be taken into consideration. An interesting demonstration here is placing my cylinder magnet 1 (Fig. 1a) on my cuboid magnet 3 (Fig. 1c). The final lowest-energy state (Fig. 2a) is found by a manual perturbation of its resting place in every direction. The final state is not the center, but close to (but still not reaching) the edge. And of course there are four of such states near the four corners, which is supported by the four-fold symmetry of such a system and also experimentally verified.

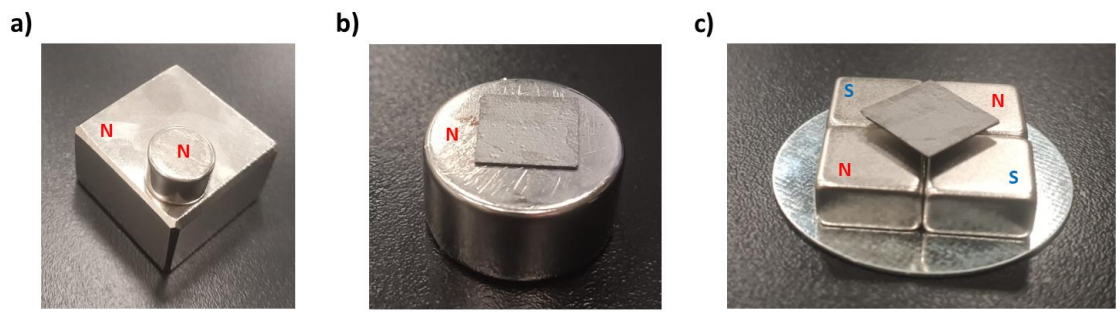

**Figure 2.** Experimental final resting place of a) A small cylinder magnet No.1 on a big cuboid magnet No.3. b) A piece of diamagnetic pyrolytic graphite touching the surface of a big cylinder magnet No.2. c) A piece of diamagnetic pyrolytic graphite levitated over a magnet array.

Now let's consider the even more interesting situations with diamagnetism. A diamagnetic object generates ***m*** in the opposite direction of nonuniform ***B***, so it gets a negative repelling force to increase (***m*** • ***B***) away from the negative direction (negative → 0). However as diamagnetism is usually very weak, so such a force may not always affect the sample due to gravity and friction (Fig. 2b). However, the story does not end here. Still remember the gradient operator $\nabla$? A larger gradient (a faster "changing rate" of a magnetic field in the space) will instantly generate a larger magnetic force. In real world, this effect can be straightforwardly observed as a permanent magnet can "lift over" and then "throws away" a piece of diamagnetic pyrolytic graphite to the desk by a few centimeters once it is pushed near the edge. This is where the magnetic field rapidly changes its directions and strength. The result of the repulsion is to make the pyrolytic graphite away from any magnetic fields, increasing the negative (***m*** • ***B***) towards zero. This phenomenon is also experimentally observed for magnets No.2. and No.3, despite they cannot levitate the graphite near their centers due to insufficient magnetic field gradients.

In fact, this is also the reason why those magnetic levitating toys using pyrolytic graphite usually have to use a magnet array (or trail). Here in my array, four NdFeB permanent magnets ($l = d = 9$ mm, $h = 5$ mm, Fig. 2c): are fixed in alternating pattern, that is, each magnet is opposite to its face-sharing neighbor. The axial surface magnetic field at the four centers of the small magnets reads ~ 530 mT, while near the center of the array there is < 50 mT in all directions, so a horizontal magnetic force always tend to bring the graphite back to the center. In the perfect scenario, the center field should be 0 as the contributions of all magnets are symmetrically cancelled out here, and such a rapid drop of hundreds of mT within a few millimeters generates huge gradient in the *xy* plane.

In addition, above the upper surface of the array, the axial magnetic field also rapidly decreases (~50 mT/mm) because the magnetic fields are mostly confined in the proximity of nearby N/S poles. A huge gradient is also generated at the *z* axis and a strong vertical magnetic force tends to levitate the graphite. Indeed, a piece of diamagnetic pyrolytic graphite can be fully levitated in the middle of the air (~3 mm) near the center of the array. This state is a balance among the horizontal magnetic force, vertical magnetic force, and gravity. Metaphorically speaking, it lays in a bowl whose bottom and walls are made of rapidly increasing magnetic fields, it simply can go nowhere. Manually moving of the "bowl" around takes the graphite away with it, but flipping the "bowl" will just lose the catch.

## Fact #3: Magnetic torque aligns the magnetic moment of an object with the magnet field.

A real object has both translational and rotational degrees of freedom. The discussions in fact #2 about magnetic force only shows the former one. If the magnetic moment ***m*** of an object is not collinear with the magnetic field ***B***, a magnetic torque is present as ***τ*** = ***m*** × ***B***.[7] Note that × is not a simple multiplication but the cross product between vectors. Such a torque will keep trying to rotate the object until there is not any angle *θ* between them. For a small magnetized ferromagnet, the torque aligns its poles with the magnetic field of a larger permanent magnet. This is the reason why a compass can indicate the (magnetic) North/South poles of the earth, but one does not have to worry about it flying away. Here the driver is mostly a magnetic torque ***τ***, rather than a magnetic force ***F****$_{mag}$*.

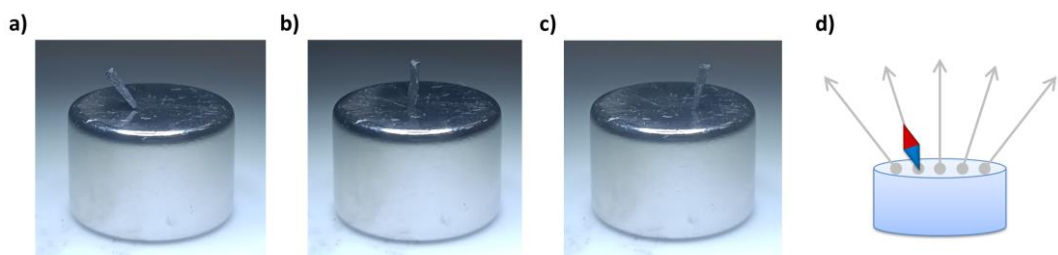

**Figure 3.** Experimental states (a-c) of a small piece of ferrite magnet attracted on the surface of a small cylinder magnet #1, along with (d) the schematic drawing of the magnetic field lines and an ideal compass.

Here I demonstrate such an effect by putting a small piece of ferrite magnet (extracted from a whiteboard pin and then purposely magnetized along the longest axis) on the surface of my NdFeB magnets (Fig. 3). Note the different angles it formed with the surface plane, which may be mistaken as so-called "half levitation". However, recalling the nonuniform distribution of the magnetic fields in Fact #1, it can simply be represented a compass that is attracted on the surface of the magnet (Fig. 3d). The real mechanism is a more complicated, and I will discuss it in more detail in Fact #5.

Another thing worth mentioning is that I am talking about "the magnetic moment *m* of an object", not "the object" itself. The magnetic moment of an object does not necessary align with the apparent physical axis/plane of a real world object, let alone an inhomogeneous mixture.

## Fact #4: A strong permanent magnet may even reset the polarity of a weak not-so-permanent magnet.

For a commercial NdFeB strong permanent magnet, its surface magnetic field can reach more than 400 mT, which is in fact larger than the coercive field of many weak not-so-permanent ferromagnetic materials including the cheap ferrite magnets. Therefore, no matter where the poles of the ferromagnetic object are at the beginning, it is not hard to reset the polarity into a new direction using a strong magnet by putting them together in the close proximity.

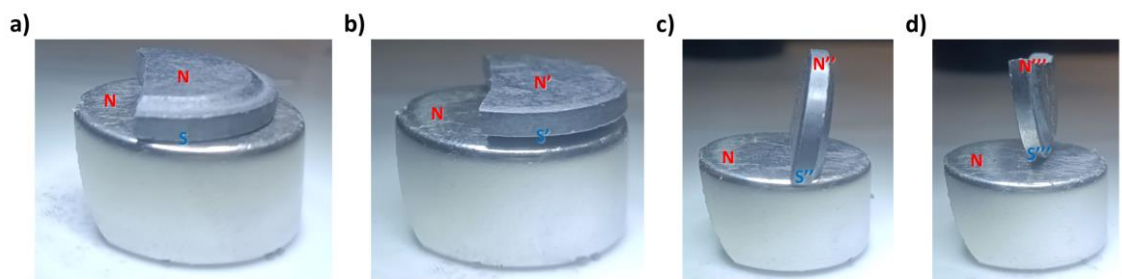

**Figure 4.** Experimental states of a large piece of ferrite magnet attracted and lying on its different surfaces on the surface of a big cylinder magnet #2

For my piece of ferrite, putting it onto the surface of a big cylinder magnet #2, followed by wiggling it around a bit, always results in attraction as the result. Checking the polarity afterwards indicates that the North/South poles of the ferrite have already been reset. Usually, one cannot observe these effects using some toy magnets, because they are all quite weak magnets compared with NdFeB. They simply cannot reset one another, unless a chosen victim is further heated to become even weaker.

## Fact #5: All these facts work together, along with other factors, generate diverse "magical" interactions even with ordinary objects

From the discussions above, it is clear that using a naive attract/repel terminology on the basis of size-less N/S of ideal magnets can never tell the whole story about a real-world magnet and its interactions with objects. To provide more take-home message, here I would like to break down the situation of some more examples of how permanent magnets interact with objects, as one may have seen from recent sources. I will ignore some entertaining ones that seem to use pastes, wires, glasses, electromagnets, static electricity and other out-of-screen factors.

**Example 5.1: "half levitation" is not levitation.**

The same piece of ferrite magnet as in Fig.3 is deliberately and carefully reset (Fact #4) so that one of the poles is at a corner. As shown in (Fig. 5a-d), it is attracted (Fact #2) to the middle of the surface of a big cylinder magnet #2. The magnetic torque (Fact #3) drives it to stand up on its corner. The local minimum of such a state is confirmed by a manual perturbation of its lifting angles in every direction, and it simply goes back.

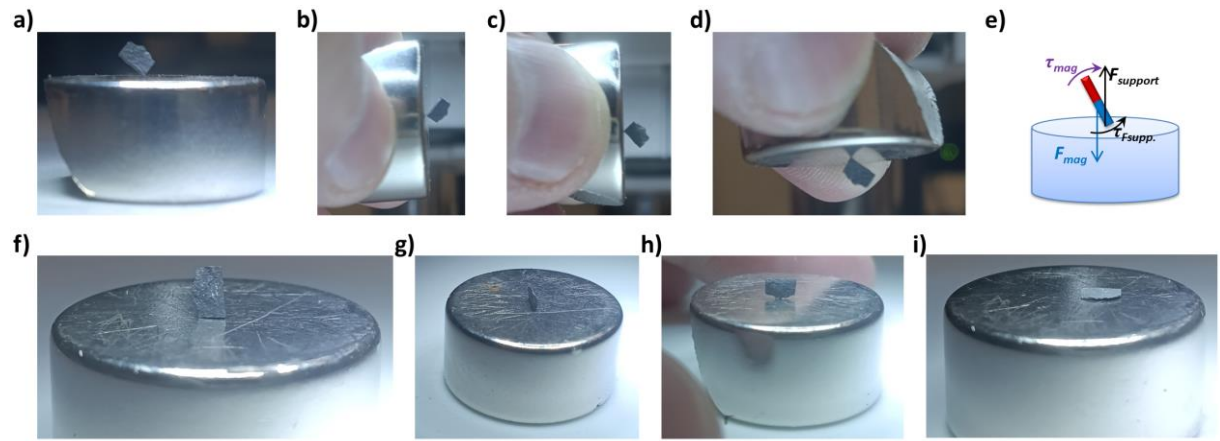

**Figure 5.** Experimental states of a small piece of ferrite magnet attracted and lying (a-e) on one of it its corners and (f-i) on different edges/surfaces on the surface of a big cylinder magnet #2. All gravities points downwards except for sub-figure (e), where gravity is ignored.

Rotating the whole system in the space does not show any visible effect of gravity, so in fact gravity can be ignored in this scenario. A simplified breakdown of the forces and torques at the mass center is shown in (Fig. 5e). It is mostly a balance among 1) attractive magnetic force, 2) repulsive supporting force, 3) aligning magnetic torque, and 4) the torque of the supporting force broken down at the mass center: $F_{support} = -F_{mag}$, $\tau_{Fsupport} = -\tau_{mag}$. Please be careful that $\tau_{mag}$ is the magnetic torque discussed in (Fact #3), not the torque generated by $F_{mag}$ (which is zero here). Alternatively, one can also breakdown the torques at the supporting point, which gives essentially equivalent results when a non-zero $\tau'_{Fmag}$ take the place of $\tau_{Fsupport}$.

Furthermore, its poles can be alternatively reset (Fact #4) into other directions, so that it can stand/lie on different edges/points/faces (Fig. 5f-i). I believe many of the readers have seen similar results in the recent sources not only from the papers,[2-5] but also from some Internet videos.[8-10] however, no superconductivity is necessary to reproduce such interactions, nor is diamagnetism.

**Example 5.2: Rise up with the torque.**

In fact, the friction $f$ between the sample and the magnet surfaces is a hidden factor for the above situations, otherwise it simply "flies" to the edges where the magnetic field (more precisely, $m \cdot B$) is larger. Here in another demonstration, the same piece of ferrite magnet is placed on a rough lab bench, with a much larger distance to a small cylinder magnet #1. When the distance decreases, the small ferrite gradually rise up with increasing lifting angles (Fig. 6a-d), until it is finally attracted to the bottom edge of the magnet (Fig. 6e).

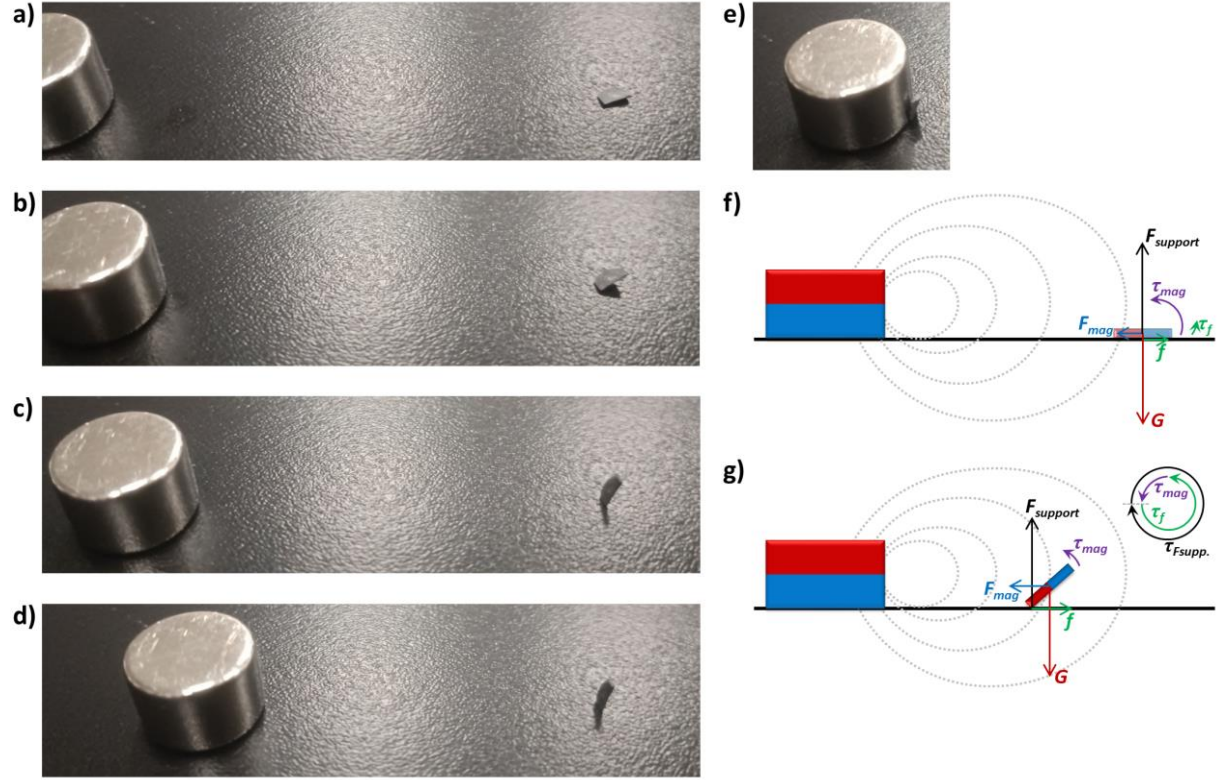

**Figure 6.** Distance-dependent lifting angles (a-d) of a small piece of ferrite magnet near a small cylinder magnet #1, (e) the final resting place, and (f-g) the simplified force analysis of its rising up.

In this case, for a hypothetical transient state at a distance (Fig. 6f), gravity $G$ and supporting force $F_{support}$ are balanced, $F_{support} = -G$. If a magnet is added to the system, the gradient of the magnetic field (more precisely, $m \cdot B$) points toward the magnet, so the object is attracted by $F_{mag}$. A friction $f$ must present to stop its movement with a balanced $f = -F_{mag}$. The magnetic field here is close to the vertical directions, so the magnetic torque $\tau_{mag}$ (along with $\tau_f$) tend to lift the farther part of the object.

If the distance is shortened and the objected rises up (Fig. 6g), $F_{support} = -G$ and $f = -F_{mag}$ still apply, despite a larger $F_{mag}$ requires a larger $f$. It is also obvious that $\tau_{Fsupport} = -(\tau_{mag} + \tau_f)$ is the key to maintain such a situation. Once $f$ exceeds the maximum static friction, the balance is lost and it goes to Fig. 6e. Once $\tau_{Fsupport}$ flips it sign by going to the other side of the mass center, the balance is also lost and it goes to Fig. 6e.

I must remind the readers that this is still a simplified analysis. Even for (Fig. 6f) the $F_{support}$ is in fact not evenly spread through the bottom surface the object, which causes a hidden $\tau_{Fsupport}$. I also do not explain why a certain lifting angle $\theta$ is corresponding to a certain distance, because it may overwhelm the readers. To make a long story short, for a given distance, the exact strength of $\tau_{Fsupport}$, $\tau_{mag}$, and $\tau_f$ are all depending on the lifting angle $\theta$, but their exact dependencies as functions of $\theta$ are not the same. A stable rise up of the object requires a solution of $\tau_{Fsupport}(\theta) = -(\tau_{mag}(\theta) + \tau_f(\theta))$ before losing its balance, which does not always exist. The exact situation is of course different from cases to cases and must be analyzed through an integration of every unit of the object, and I am only demonstrating one of the possibilities. Anyway, no superconductivity is necessary to reproduce such interactions, nor is diamagnetism.

**Example 5.3: No friction, no obstacles.**

If there is no friction, or at least it is minimized to an extent, a breakdown of the scenario indicates very limited final results. For a ferromagnetic/paramagnetic object, the energy minimum due to attraction is near the edge (not the center, Fact #1) of a real-world permanent magnet (Fig. 7a). Similar demonstrations are easily found on the Internet, but I cannot identify the original one. Here I demonstrate such a situation using a small piece of ferromagnetic Nickel metal attached to some wooded toothpicks so that it can float on the water to minimize friction (Fig. 7b). No matter where it is placed at the beginning, it is always attracted toward and stabilized near one of the edge of the magnet (Fact #2). It also comes and goes with the magnet, because this is the energy minimum (Fig. 7c).

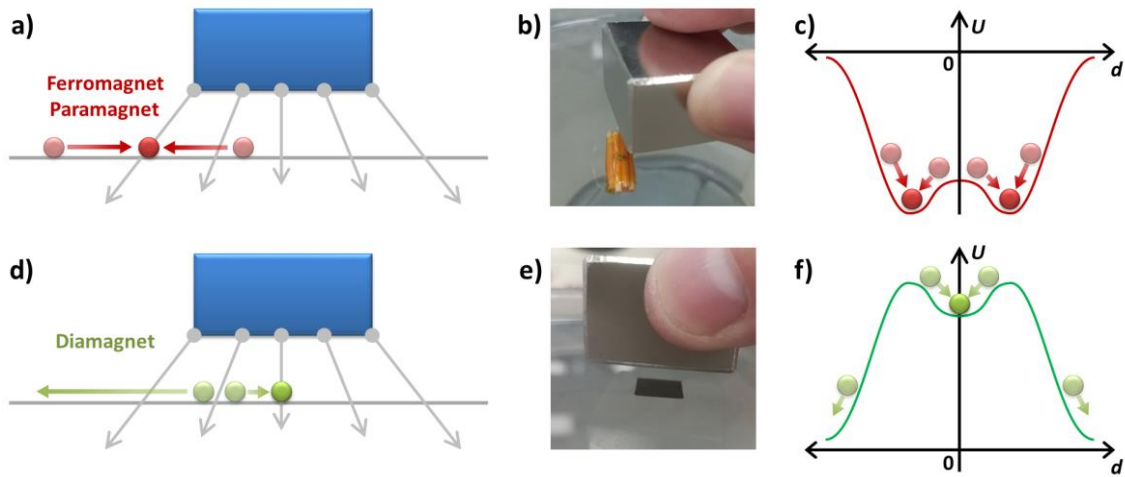

**Figure 7.** Experimental states and schematic distance-dependence of the magnetic potential energy of (a-c) a ferromagnetic Nickel metal attached to some wooded toothpicks and floating on the water and (d-f) a diamagnetic pyrolytic graphite floating on the water and their final states near a big cuboid magnet No.3.

For a diamagnetic object, however, the energy minima due to repulsion can have two results: If the object is placed close to the center at the beginning, it is repelled, but the repulsion is stronger at the edge than it is at the center. (Fig. 7d, Fact #2) So with a careful try, one can sometimes stabilize a small piece of diamagnetic pyrolytic graphite near the center of a big cuboid magnet No.3 (Fig. 7e). This is a local minimum (Fig. 7f). On the other hand, if the object is placed close to the edge or even further at the beginning, it is repelled far away. In a distance, there is basically no magnetic field and the potential energy is minimized toward zero (Fig. 7f, experimental results are not shown). Again, no superconductivity is necessary to reproduce such interactions. In fact if a superconductor is played in this way, it should simply be "pinned" wherever it is at the first place.

## Concluding Remarks

From the discussions above, it is demonstrated how complicated a real-world permanent magnet can interact with even ordinary objects. The diversity of these interactions comes from many factors including (bnt not limited to): 1) the non-ideal distribution of the magnetic fields near the surface, 2) the gradient of a magnetic field that moves object, 3) the overlooked magnetic torque that aligns object, 4) the possible flipping of the polarity, and 5) the combination with other forces/torques such as supporting force, friction and gravity. Some of these interactions should have been taught in a middle school, at least in a high school. However, our education system seems fail to do so, at least not for everyone.

Nevertheless, one can still play around with some pieces of magnets in one's spare time, like a curious child or a responsible scientist would have done, to be amazed by its diversity and beauty. I would like to point out again that, the most tricky part of most of these demonstrations, in fact, is to choose a weak-enough object, otherwise it may simply "slip/jump/fly/teleport" to the edge of the strong magnet due to the gradient. I hope these discussions are helpful to those who want to think deeper about magnetic mechanism that leads to these diverse interactions, especially when one or more of the factors may be unknown or overlooked. Finally, I wish that the next time when some ordinary magnetic tricks are claimed to support extraordinary magical findings, a real scientist could think twice before spending the taxpayers' money.

## Acknowledgments

This work is not supported by any external funding other than the author's own curiosity and responsibility. The author thanks the Central Lab of the School of Chemistry, Sun Yat-Sen University for some lab benches to play around with the magnets.